**Unraveling the role of the magnetic anisotropy on thermoelectric response: a theoretical and experimental approach**

M. A. Corrêa[1,2,a], M. Gamino[1], A. S. de Melo[1], M. V. P. Lopes[1], J. G. S. Santos[1], A. L. R. Souza[1], S. A. N. França Junior[1], A. Ferreira[2], S. Lanceros-Méndez[2,3,4], F. Vaz[2] and F. Bohn[1]

[1] *Departamento de Física, Universidade Federal do Rio Grande do Norte, 59078-900 Natal, RN, Brazil*

[2] *Centro de Física, Universidade do Minho, 4710-057 Braga, Portugal*

[3] *BCMaterials, Basque Center for Materials, Applications and Nanostructures, UPV/EHU Science Park, 48940 Leioa, Spain*

[4] *IKERBASQUE, Basque Foundation for Science, E-48009 Bilbao, Spain*

---

[a] marciocorrea@fisica.ufrn.br

Abstract

Magnetic anisotropies have key role to taylor magnetic behavior in ferromagnetic systems. Further, they are also essential elements to manipulate the thermoelectric response in Anomalous Nernst (ANE) and Longitudinal Spin Seebeck systems (LSSE). We propose here a theoretical approach and explore the role of magnetic anisotropies on the magnetization and thermoelectric response of noninteracting multidomain ferromagnetic systems. The magnetic behavior and the thermoelectric curves are calculated from a modified Stoner Wohlfarth model for an isotropic system, a uniaxial magnetic one, as well as for a system having a mixture of uniaxial and cubic magnetocrystalline magnetic anisotropies. It is verified remarkable modifications of the magnetic behavior with the anisotropy and it is shown that the thermoelectric response is strongly affected by these changes. Further, the fingerprints of the energy contributions to the thermoelectric response are disclosed. To test the robustness of our theoretical approach, we engineer films having the specific magnetic properties and compare directly experimental data with theoretical results. Thus, experimental evidence is provided to confirm the validity of our theoretical approach. The results go beyond the traditional reports focusing on magnetically saturated films and show how the thermoelectric effect behaves during the whole magnetization curve. Our findings reveal a promising way to explore the ANE and LSSE effects as a powerful tool to study magnetic anisotropies, as well as to employ systems with magnetic anisotropy as sensing or elements in technological applications.

I. INTRODUCTION

Thermoelectric effects driven by spin currents thermally activated have attracted increasing interest not just due to their relevance in the context of fundamental physics of phenomena associated with spin caloritronics [1–5], but also due to their potential relevance in a wide variety of technological applications [6–8]. Remarkably, the Anomalous Nernst Effect (ANE) [9–12] and the Longitudinal Spin Seebeck Effect (LSSE) [7,13–17] arise as the most studied effects in the field.

ANE and LSSE are observed when a magnetic material is simultaneously submitted to a temperature gradient $\nabla T$ and a magnetic field $\vec{H}$ perpendicular to each other. However, while ANE appears exclusively in ferromagnetic (FM) metals, LSSE is often disclosed for bilayers consisting of an insulating ferrimagnetic (FMI) or FM layer capped by non-magnetic metal (NM) with high spin-orbit coupling [7,13–17].

It is worth highlighting that, in ferromagnetic metallic bilayer heterostructures, the thermoelectric response of the whole system is a result of the combination of both effects, ANE and LSSE. Hence, to circumvent this fact and suppress any contribution associated with ANE or thermomagnetic effects and to , thus obtain a pure LSSE signal, systems are commonly engineered, taking into consideration a ferrimagnetic insulator, for instance the well-known $Y_3Fe_5O_{12}$ (YIG) material, to compose the bilayers heterostructure [18–21]. Indeed, the magnetic and electrical nature of YIG suggests the YIG/NM bilayers as natural candidates for investigations addressing the LSSE.

Despite the suitability of the YIG/NM bilayers for such goal, the experimental procedures needed to induce the ferrimagnetic phase in YIG thin films often involve several steps, including reactive magnetron sputtering and/or annealing, the latter typically performed in a controlled atmosphere [22]. In addition, the magnetic behavior of the YIG/NM systems is commonly found to be isotropic in the film-plane. These features make the achievement of YIG/NM bilayers combining specific structural and magnetic properties a complex task.

Given these issues raised by YIG/NM bilayers, a different path would be highly convenient. Thus, we bring to light that ferromagnetic films, such as Co-rich alloys, may reveal magnetic field dependent thermoelectric responses with interesting characteristics [23]. Through the appropriate selection of substrates and/or experimental deposition parameters, it is possible to induce/control the magnetic anisotropies of the FM system. Consequently, the thermoelectric response may be modified by both, the magnetic anisotropy and the orientation of the external magnetic field [23].

While the thermoelectric effects have been widely investigated from the theoretical perspective, the role of the magnetic anisotropy on thermoelectric effects has rarely been addressed, despite of its scientific and technological potential relevance [24]. Recently, the modifications of the magnetic properties and thermoelectric effects, as ANE, have been explored in flexible magnetostrictive $Co_{40}Fe_{40}B_{20}$ thin films with induced uniaxial magnetic anisotropy and under external stress [23]. Moreover, the influence of the cubic magnetocrystalline anisotropy of $Co_{58.73}Fe_{27.83}Al_{13.44}$ Heusler alloys, grown onto substrates with distinct orientations, on thermoelectric effects has been studied by the LSSE [25]. For both situations, the results corroborate that the thermoelectric curves are strongly dependent on the magnitude and orientation of the magnetic field with respect to the anisotropies [23–25]. In addition, the thermoelectric measurements may be yet used to probe the magnetic anisotropy, as well as to identify the field ranges in which the anisotropic and Zeeman energy terms represent the major contribution to the magnetic free energy density governing the magnetization process [25].

In this work, we propose a theoretical approach from a modified Stoner Wohlfarth model [26] and explore the role of magnetic anisotropies on the magnetic and thermoelectric responses of noninteracting multidomain ferromagnetic systems. The magnetic behavior has been calculated together with the thermoelectric response for an isotropic system, a uniaxial magnetic one, as well as for a system having a mixture of uniaxial and cubic magnetocrystalline magnetic anisotropies. Modifications of the magnetic behavior with the anisotropy have been verified and it is shown that the thermoelectric response is strongly affected by these changes. To test the robustness of the developed theoretical approach, films have been engineered having the specific magnetic properties and a direct comparison between experimental data and theoretical results has been carried out. The results go beyond the traditional reports focusing on magnetically saturated films and show how the behavior of the thermoelectric effect during the whole magnetization curve.

## II. THEORETICAL APPROACH

### A. Thermoelectric effects

The Anomalous Nernst Effect corresponds to the generation of a voltage by applying a temperature gradient to a ferromagnetic metal or a semiconductor [9,10]. The electrical field associated with the ANE can be expressed as

$$\vec{E}_{ANE} = -\lambda_N \mu_\circ (\vec{m} \times \nabla T), \qquad (1)$$

in which $\lambda_N$ is the anomalous Nernst coefficient, $\mu_\circ$ is the vacuum magnetic permeability, $\vec{m}$ is the magnetization vector, and $\nabla T$ is the temperature gradient.

The Spin Seebeck Effect (SSE), in turn, consists in the generation of a voltage by applying a temperature gradient to FM/NM or FIM/NM heterostructures [7,27]. The electrical field due to the SSE generated in the non-magnetic metal is described by

$$\vec{E}_{SSE} = -\lambda_{SSE}(\vec{\sigma} \times \nabla T). \tag{2}$$

Here, $\lambda_{SSE}$ is the spin Seebeck coefficient, $\vec{\sigma}$ is the spin polarization whose orientation is given by the magnetization direction, and $\nabla T$ represents the temperature gradient.

From Eqs. (1) and (2), it can be noticed the remarkable similarity of the effects coming also from the experimental configuration, where $\nabla T$ and $\vec{H}$ are perpendicular to each other, as represented in Fig. 1. In the present case, it is assumed that $\nabla T$ is perpendicular to the plane of the film, while $\vec{H}$ remains in the plane of the film, as depicted in Fig. 1. For the calculations, the thermoelectric voltage *V* detected at the electrical contacts at the ends of the main axis of the film is given by:

$$V = -\int_0^L \vec{E} \cdot \vec{dl}, \tag{3}$$

the limits of integration being related to the distance between the electrical contacts used in the experiment. Here we assume *y* as the direction for the detection of *V*, as shown in Fig. 1.

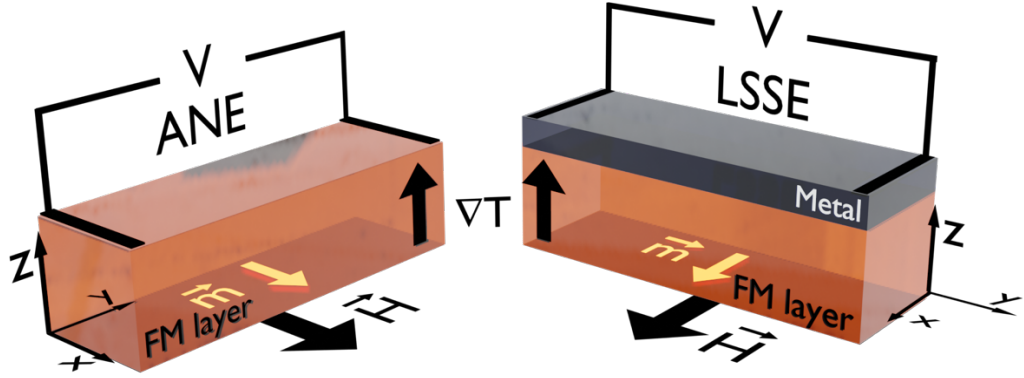

FIG. 1. Schematic representation of (Left) ANE and (Right) LSSE experiments. For the LSSE system, the non-magnetic metal shall have high-spin orbit coupling constant, hence yielding the conversion of the spin current to charge current through the Inverse Spin Hall Effect. In our calculation, it is assumed *y* as the direction to the *V* detection.

The SSE voltage is measured by means of the Inverse Spin Hall Effect (ISHE) [28]. The relation between the spin current $J_s$, generated from the temperature gradient, and the charge current $J_c$, produced in the NM metal layer, is given by:

$$\vec{J}_c = \theta_{SH}\left(\frac{2e}{\hbar}\right)(\vec{J}_s \times \hat{\sigma}), \quad (4)$$

in which $\hat{\sigma}$ is the spin-polarization direction and $\theta_{SH}$ is the spin Hall angle that measures the efficiency of the spin-charge conversion.

B. Magnetic free energy density

From the magnetic free energy density $\xi$ describing the system with a given anisotropy configuration [29], a routine for the energy minimization provides the values of the equilibrium angles of the magnetization at each magnetic field $\vec{H}$. Then, once the magnetization response is obtained for a specific field orientation, the thermoelectric voltage *V* can be obtained. This numerical procedure enables to achieve the dependence of the thermoelectric voltage for systems with distinct anisotropy configurations.

To proceed with the magnetic and thermoelectrical calculations, the film defining the magnetic system is first described. Here, a theoretical system is considered that is divided into 50 noninteracting magnetic domains, what allows to represent a sample having magnetic anisotropy dispersion. The magnetization of the whole system is then obtained averaging the magnetization of the domains. The magnetic free energy density for each domain is based on a modified Stoner-Wohlfarth model [30,31], in which the magnetic anisotropy is considered. Here, we have focus on systems having uniaxial and cubic magnetocrystalline anisotropies, as well as on an isotropic one. For the sake of simplicity, the magnetization of each domain is considered in the plane of the film, and consequently, the free energy density can be written by

$$\xi_i = -m_{si} \, H \cos(\varphi_H - \varphi_{mi}) - k_{ui} \, \sin^2(\varphi_{ki} - \varphi_{mi}) - \frac{1}{4}\xi_{c1i}. \tag{5}$$

The first term is the Zeeman interaction, the second one describes the uniaxial magnetic anisotropy, and the last one is related to the cubic magnetocrystalline anisotropy. For the numerical calculations, $\theta_{mi} = 90°$ and $\varphi_{mi}$ are defined as the equilibrium angles of the magnetization for each domain at a given magnetic field $\vec{H}$. At the same time, $\theta_{mi}$ is the angle of the magnetization vector with respect to the $z$ axis, while $\varphi_{mi}$ is the angle between the projection of the $\vec{m}_i$ in the $xy$ plane with respect to the $x$ axis, as indicated in Fig. 2. Similarly, $\theta_H$ and $\varphi_H$ are defined as the angles describing the orientation of the magnetic field $\vec{H}$. In addition, $\widehat{m}_i$ is the unit vector of the magnetization, $m_{si}$ is the saturation magnetization, and $k_{ui} = \frac{m_{si} h_{ki}}{2}$, where $h_{ki}$ is the anisotropy field related to the uniaxial magnetic anisotropy in the $\hat{u}_{ki}$ direction for each domain, defined by $\theta_{ki}$ and $\varphi_{ki}$. For all quantities, the subscript *i* indicate that they are associated to the *i-th* magnetic domain composing the system. Finally, $\xi_{c1i}$ describes the in-plane component of cubic magnetocrystalline anisotropy, which can be written as

$$\xi_{c1i} = k_{c1i}(\alpha_{1i}^2\alpha_{2i}^2 + \alpha_{1i}^2\alpha_{3i}^2 + \alpha_{2i}^2\alpha_{3i}^2), \tag{6}$$

with

$$\alpha_{1i} = \cos\varphi_{mi} \sin\theta_{mi},$$

$$\alpha_{2i} = \sin\varphi_{mi} \sin\theta_{mi}, \tag{7}$$

$$\alpha_{3i} = \cos\theta_{mi}.$$

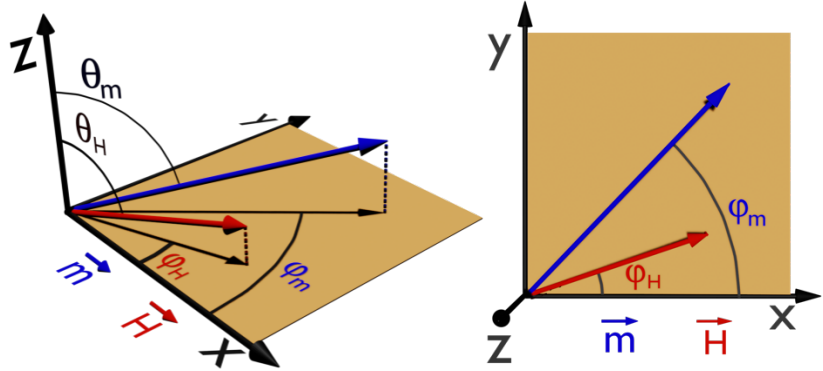

FIG. 2. Schematic diagram of the theoretical system and definitions of the vectors employed for the numerical calculation. Here we consider m as the magnetization vector, whose orientation for each given magnetic field value is set by $\theta_m$ and $\varphi_m$, the equilibrium angles with respect to the z and x axes, respectively. The $\vec{H}$ corresponds to the external magnetic field vector, described by $\theta_H$ e $\varphi_H$. In particular, the uniaxial anisotropy vector (not shown here) is in the film-plane, i. e. $\theta_{ki} = 90°$.

Looking at Eq. (5), a system with uniaxial magnetic anisotropy is represented taking $k_{ui} \neq 0$ and $k_{c1i} = 0$. From our concern here, we may mention that taking $k_{ui} \neq 0$ and $k_{c1i} \neq 0$, a system is described having both uniaxial and cubic magnetocrystalline anisotropies, while by considering $k_{ui} \neq 0$, $\varphi_m = \varphi_H$ and $k_{c1i} = 0$, a system is described in which the magnetization is aligned to the magnetic field, i.e. having isotropy in the film-plane.

1. *Magnetic system with isotropy in the film-plane*

Figure 3 presents the magnetic and thermoelectric responses for a magnetic system having magnetic isotropy in the plane of the film. As previously mentioned, it is assumed $k_{ui} \neq 0$, $\varphi_{ki} = \varphi_H$ and $k_{c1i} = 0$ to describe the magnetic free energy density of this case of study. To our numerical calculations, a system is considered consisting of a single magnetic domain with the following general parameters: $m_{si} = 143$ emu/cm³ and $h_{ki} = 30$ Oe, with $\theta_{ki} = 90°$, $\varphi_{ki} = \varphi_H$, and $\theta_H = 90°$. Notice that the use of the multidomain structure is not required for this case. For the thermoelectric calculation, it is assumed Δ*T* = 30 K and an effective thermoelectric coefficient of $\lambda_{eff} = 1.3 \times 10^{-5}$ V/K [23]. Here $\lambda_{eff}$ brings the contributions of $\lambda_N$ and $\lambda_{SSE}$ of the system. For a system in which the LSSE is also present, such as YIG/NM, it is assumed $\lambda_{eff} = \lambda_{SSE}$.

For sake of simplicity, the results are addressed in terms of the normalized magnetization and thermoelectric voltage values. To this end, the quantities $m/m_s$ and $V/V_{max}$ are defined, where $m_s$ and $V_{max}$ are the maximum values obtained for the magnetization and the thermoelectric voltage, respectively, when $\varphi_H = 0°$.

Figure 3(a) shows the magnetization behavior as a function of the field for distinct $\varphi_H$ values. As expected, all the curves have squared shapes, with coercive field of 30 Oe, being similar irrespective of the field orientation given by $\varphi_H$, a fact due to the constant alignment between the $\vec{m}_{si}$ and $\vec{H}$ for isotropic systems.

Figure 3(b) in turn shows the thermoelectric response for such isotropic magnetic system. The shape of the curves mirrors the magnetization loops, even with varying $\varphi_H$. However, it is observed a significant reduction in the amplitude of the curve as $\varphi_H$ increases. This feature is directly related to the decrease of the component of the induced electric field along the *V* detection direction defined by the electrical contacts.

Figure 3(c) presents the angular dependence of the $V/V_{max}$ for constant magnetic field value. Here we show the curves for two magnetic field values, 300 Oe and 2000 Oe. For both situations, the sample is magnetically saturated, and the signal response has a cosine-shaped form, being proportional to $\cos \varphi_H$.

Finally, it is worth emphasizing that the results for the isotropic magnetic system are in agreement with the ones found in previous reports for ferrimagnetic materials, such as the YIG/Pt bilayers, samples that are widely explored in investigations addressing LSSE effects [7,32].

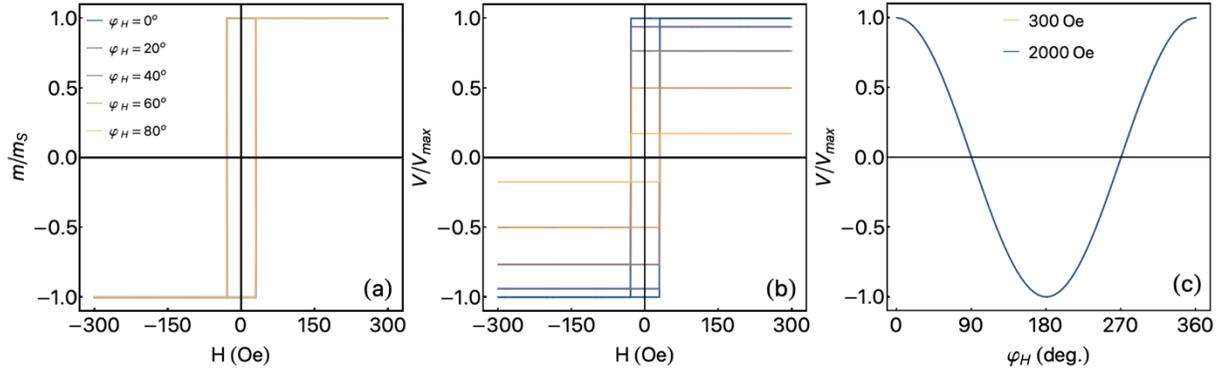

FIG. 3. (a) Normalized magnetization curves and (b) thermoelectric voltage curves at distinct $\varphi_H$ for a system having isotropic magnetic features in the film-plane. For the calculations, the following parameters are considered: $m_{si} = 143$ emu/cm³, $h_{ki} = 30$ Oe, $\theta_{ki} = 90°$, $\varphi_{ki} = \varphi_H$, $\theta_H = 90°$, $\varphi_H$ ranging from 0° up to 90°, $\lambda_{eff} = 1.3 \times 10^{-5}$ V/K and $\Delta T$ = 30 K. (c) $V/V_{max}$, at $H = 300$ Oe and 2000 Oe and with $\Delta T$ = 30 K, as a function of $\varphi_H$.

2. *Magnetic system with uniaxial magnetic anisotropy*

For a magnetic system with uniaxial magnetic anisotropy, it is assumed $k_{ui} \neq 0$ and $k_{c1i} = 0$. Specifically, a system is considered with the following parameters: $m_{si} = 1330$ emu/cm³, $h_{ki} = 100$ Oe, $\theta_{ki} = 90°$, $\theta_H = 90°$, and $\lambda_{eff} = 1.3 \times 10^{-5}$ V/K. In addition, it is now taken into account the multidomain structure in order to mimic the magnetic anisotropy dispersion. The uniaxial anisotropy direction is first set at $\varphi_{ki} = 90°$, with linear dispersion of 5°. This procedure enables to represent a ferromagnetic system in which the uniaxial magnetic anisotropy is affected/dispersed by, for instance, the stress stored in a film during deposition.

Figure 4 shows the evolution of the magnetic and thermoelectric responses with $\varphi_H$ for a magnetic system with uniaxial magnetic anisotropy, in which the easy magnetization axis lies in the plane of the film. For the thermoelectric calculations, it is also assumed $\Delta T$ = 30 K.

Figure 4(a) shows the magnetization curves for distinct values of $\varphi_H$. From a general point of view, it is verified the well-known behavior with a near-squared shape for the curves calculated at $\varphi_H = 80°$, revealing an easy magnetization axis located nearby this orientation. For the $\varphi_H = 0°$, the form of the magnetization curve directly allows to infer a hard magnetization axis, corroborating the behavior for an uniaxial magnetic system when the field is perpendicular to the easy axis. Remarkably, it is observed a small coercive field for $\varphi_H = 0°$, a feature arising from the magnetic domains having anisotropy dispersion. In this sense, the presented theoretical approach considering multidomains appears as an appropriate tool to describe details of the magnetic response of the systems.

Considering the thermoelectric behavior, presented in Fig. 4(b), the shape of the curves is not strongly modified as the $\varphi_H$ increases. For $\varphi_H = 0°$, the thermoelectric curve mirrors the one observed for the magnetization. This feature is expected, since the induced electrical field is along the $V$ detection direction given by the electrical contacts for this configuration. However, although the shape of the curves remains unchanged even with increasing $\varphi_H$ (from 0 to 90°), it is observed a decrease in the $V/V_{max}$ value at maximum field.

Figure 4(c) depicts the dependence of $V/V_{max}$ with $\varphi_H$ for a constant magnetic field. Here, an interesting feature is observed with respect to the shape of the curves.

For an external magnetic field of 2000 Oe, it is observed that the response of $V/V_{max}$ as a function of $\varphi_H$ is perfectly described by a cosine, as expected according to theory for a magnetically saturated sample. For an external field of 300 Oe in turn, it is verified a change in the profile of the curve and the sample is magnetically saturated at this smaller field value, an intriguing fact at a first glance.

At 2000 Oe, the field value is much above the saturation and anisotropy fields. As a consequence, the Zeeman energy is, at least, one order of magnitude higher than the uniaxial anisotropy energy, representing the major contribution to the magnetization dynamics. Hence, the Zeeman energy defines the magnetization direction, which follows the orientation of the field, leading to a dependence of $V/V_{max}$ with $\cos \varphi_H$.

The most striking finding here resides in the fact that the deviation from the cosine shape is found when the field is 300 Oe. This distortion can be interpreted in terms of the competition of two energy contributions. Specifically, at a field value right above the saturation, the Zeeman energy and the uniaxial magnetic anisotropy energy are of the same order of magnitude, in a sense that the response of the whole system is a result of the balance of both of terms. Given that, small deviations of the magnetization are identified through the thermoelectric experiment.

Remarkably, although the magnetometric techniques are not efficient enough to probe small deviation of the magnetization at saturate states, in which the field is right above the saturation field, the ANE or LSSE experiments provide an increased sensitivity for such task.

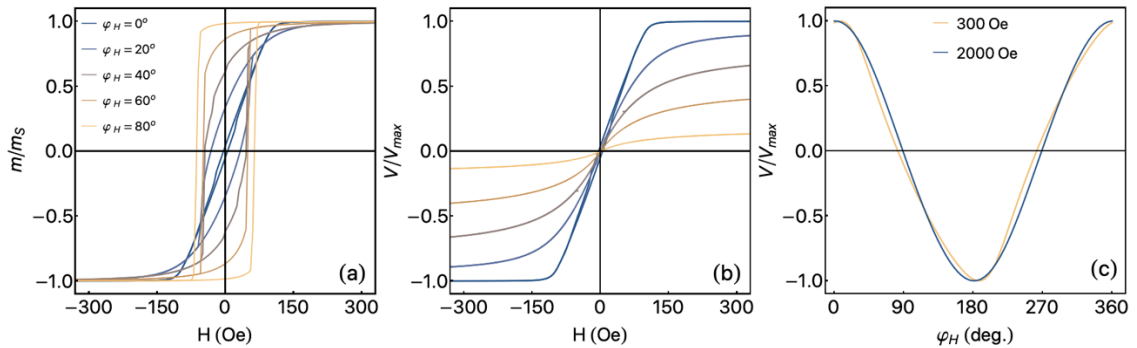

FIG. 4. (a) Normalized magnetization curves and (b) thermoelectric voltage curves for a system with uniaxial magnetic anisotropy with $\varphi_{ki} = 90°$. For the calculations, the following parameters are considered: $m_{si} = 1330$ emu/cm³, $h_{ki} = 100$ Oe, $\theta_{ki} = 90°$, $\theta_H = 90°$, $\varphi_H$ ranging from 0° up to 90°, $\lambda_{eff} = 1.3 \times 10^{-5}$ V/K and $\Delta T = 30$ K. (c) $V/V_{max}$, at $H = 300$ Oe and 2000 Oe and with $\Delta T = 30$ K, as a function of $\varphi_H$.

To explore the modification on the thermoelectric curves when the uniaxial anisotropy direction is modified, now it is considered $\varphi_{ki} = 0°$, with a linear dispersion of $5°$, as used in the previous case. Fig. 5(a) shows the magnetization curves, allowing to observe the quasi-static magnetic behavior for a uniaxial system with in-plane magnetic anisotropy. Instead of the previous situation ($\varphi_{ki} = 90°$), the magnetization curves start from a squared shape for $\varphi_H = 0°$. Besides, the modification of $\varphi_H$ shows similar curves to the ones observed in the previous case, in which $\varphi_{ki} = 90°$. On the other hand, remarkable modifications are observed in the thermoelectric curves, as shown in Fig. 5(b). While the shape of the curves is unchanged in the previous situation with varying $\varphi_H$, here the thermoelectric curves present a strong dependence on $\varphi_H$ values. For $\varphi_H = 0°$ the thermoelectric curves mirror the magnetization curves, with similar coercive field values, as expected, since the electric field is aligned with the electrical contacts. However, as $\varphi_H$ increases, a decrease in the thermoelectrical intensity at high fields is observed, keeping unchanged the maximum values reached by the thermoelectric voltage when $\vec{H}$ is close to $h_{ki}$. The shape of the curves is a result of the competition between the Zeeman and the uniaxial anisotropy, what is expected given that the sample is not saturated. For the curves at higher $\varphi_H$ values and high $\vec{H}$ intensities, the magnetization is aligned with the magnetic field direction. However, as the magnetic field intensity decreases, the uniaxial magnetic anisotropy leads to a rotation of the magnetization and, consequently, of the electric field in the LSSE or ANE experiments. This feature leads to an increase in the thermoelectric voltage for fields near to $h_{ki}$ values.

Once again, for the $V/V_{max}$ as a function of the $\varphi_H$ curves (Fig. 5 (c)), it is observed a slight distortion on the cosine curve for an external field of 300 Oe, a feature associated to the competition of the Zeeman and uniaxial anisotropy. However, the cosine shape is observed for the curves in which the field is 2000 Oe, in agreement with the previous discussions.

Considering the remarkable modification in the thermoelectric curves observed when the $\varphi_{ki}$ values are changed (Figs. 4 and 5), the magnetic and thermoelectric curves have been calculated for a fixed $\varphi_H$, and the $\varphi_{ki}$ varying from 0° up to 180°, these curves being depicted and discussed in Fig. 1S of the Supplementary Material.

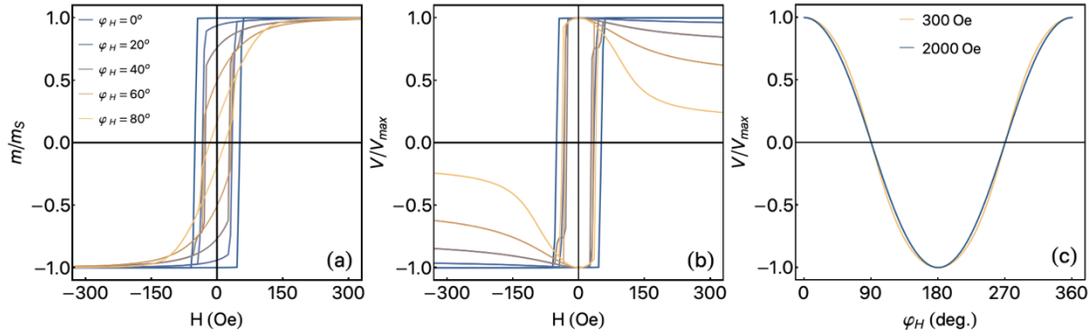

FIG. 5. (a) Normalized magnetization curves and (b) Normalized thermoelectric voltage curves for a system with uniaxial magnetic anisotropy with $\varphi_{ki} = 0°$. For the calculations, the following parameters have been considered: $m_{si} = 1330$ emu/cm³, $h_{ki} = 100$ Oe, $\theta_{ki} = 90°$, $\theta_H = 90°$, $\varphi_H$ ranging from 0° up to 90°, $\lambda_{eff} = 1.3 \times 10^{-5}$ V/K and $\Delta T$ = 30 K. (c) $V/V_{max}$, at $H = 300$ Oe and 2000 Oe and with $\Delta T$ = 30 K, as a function of $\varphi_H$.

3. *Magnetic system with uniaxial and cubic magnetocrystalline anisotropies*

A more complex magnetic anisotropy configuration was further explored. In the following, a magnetic system having both uniaxial and cubic magnetocrystalline anisotropies is addressed. Such system discloses very interesting magnetization curves and thermoelectric responses, with potential for technological applications, such as in magnetic logic keys.

Specifically, a system is considered with the parameters previously employed for the uniaxial one, $m_{si} = 1330$ emu/cm³, $h_{ki} = 100$ Oe, $\theta_{ki} = 90°$, $\theta_H = 90°$, and $\lambda_{eff} = 1.3 \times 10^{-5}$ V/K, with the additional quantity of $k_{c1i} = 2.8 k_{ui}$ associated with the cubic magnetocrystalline anisotropy.

The multidomain structure is also kept to represent the magnetic anisotropy dispersion, by setting the uniaxial anisotropy direction at $\varphi_{ki} = 90°$, with a linear dispersion of 5°.

Figure 6 presents the numerical calculations for the magnetization curves and thermoelectric responses for $\varphi_{ki} = 0°$, $\varphi_{ki} = 45°$, and $\varphi_{ki} = 90°$.

Considering $\varphi_{ki} = 0°$, Figs. 6 (a.I) and (a.II) depict the magnetization and thermoelectric behavior as a function of the external magnetic field for distinct $\varphi_H$ values. The addition of the cubic magnetocrystalline anisotropy yields, besides the easy and hard magnetization axes, the intermediate one at $\varphi=45°$. The emergence of such configuration of magnetic anisotropy and the competition of anisotropy axes lead to magnetization curves with interesting behavior, in particular for high $\varphi_H$ values. Given that the uniaxial anisotropy axis is along the V detection direction, the richness of details found in the magnetization curves is not observed in the thermoelectric ones.

However, when $\varphi_{ki} \neq 0°$, the refined structure of the magnetization curves, full of details associated to the magnetization process, is also identified in the thermoelectric curves. For $\varphi_{ki} = 45°$, changes in both, magnetization and thermoelectric curves are observed as the $\varphi_H$ is modified (see Figs. 6 (b.I) and (b.II)). For $\varphi_H$ values smaller than 45°, the thermoelectric curves present similar shape to the one observed on the magnetization curves. On the other hand, for $\varphi_H$ values higher than $\varphi_{ki} = 45°$, the shapes of the thermoelectric and magnetic curves disclose fundamental differences, with a decrease in $V/V_{max}$. Taking into account $\varphi_{ki} = 90°$ (Fig. 6 (c.I) and (c.II)), the mirror of the magnetization curves and thermoelectric one is verified, for $\varphi_H = 0°$ and $\varphi_H = 20°$ field directions. On the other hand, a remarkable modification on the thermoelectric curves is verified for $\varphi_H$ above 45°. This behavior reflects the modifications observed in the previous situations. From a general point of view, important modifications are observed in the shape of the thermoelectric curves when $\varphi_H$ is higher than 45°, irrespectively of the $\varphi_{ki}$ direction. This behavior corroborates previous results observed in magnetostrictive systems, in which the effective magnetic anisotropy has been modified by stress applications [23].

It is worth remarking that the addition of the cubic magnetocrystalline to an uniaxial anisotropy axis (keeping constant the anisotropy parameters considered in the previous sections and assuming $k_{c1i} = 2.8k_{ui}$) describes a system with harder magnetic properties, if compared with the ones found in Fig. 3(a), 4(a) and 5(a). This fact becomes evident when the coercive fields are observed in Fig. 6(a.I, b.I and c.I), as well as through the fact that most of the curves in Fig. 6 do not achieve the magnetic saturation at 300 Oe.

Such behavior is reflected in the dependence of $V/V_{max}$ with $\varphi_H$, as presented in Fig. 6 (d.I) for a system with $\varphi_{ki} = 0°$. Specifically, distortions in the curves are evidenced when the external magnetic field is 300 Oe. As previously discussed, deviations in the expected cosine form are attributed to a competition between the contributions of the Zeeman energy and anisotropy terms. For a field of 2000 Oe in turn, the Zeeman energy becomes at least one order of magnitude greater than the uniaxial and cubic anisotropies ones, having the major contribution to the magnetization dynamics. In this case, the system is magnetically saturated, and the cosine shape is recovered.

To highlight this feature, Fig. S2 in the Supplementary material presents calculations of the magnetization and thermoelectric response for a field of $\pm 2000$ Oe. To highlight the influence of the magnetic state on the thermoelectric response, Fig. 6(d.II) brings the evolution of the $V/V_{max}$, at 300 Oe, with $\varphi_H$ for distinct $\varphi_{ki}$ values. From this, the competition of the anisotropies and Zeeman energies in this complex magnetic system can be disclosed.

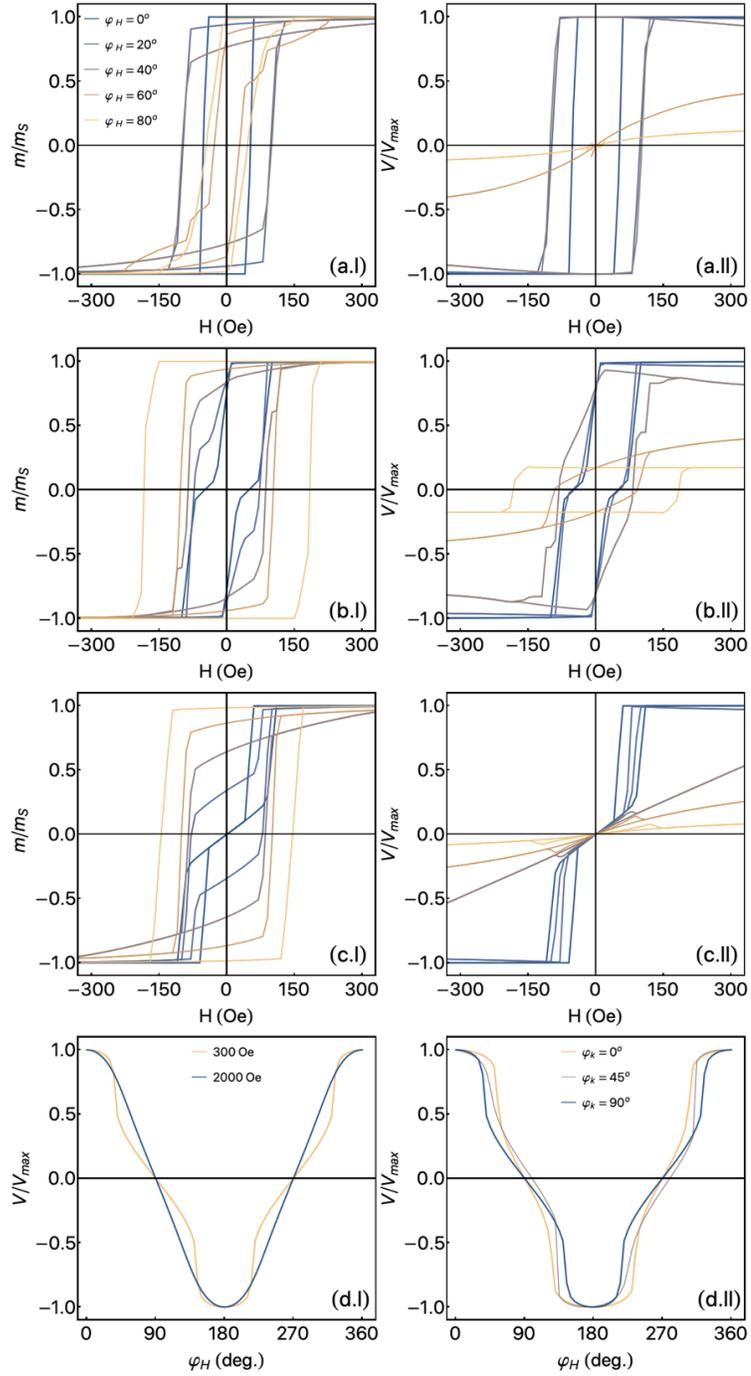

FIG. 6. Numerical calculation for a system with uniaxial and cubic magnetocrystalline anisotropies. Normalized magnetization, at selected field orientations, for a system with (a.I) $\varphi_{ki} = 0°$, (b.I) $\varphi_{ki} = 45°$ and (c.I) $\varphi_{ki} = 90°$. Thermoelectric curves for a system with (a.II) $\varphi_{ki} = 0°$, (b.II) $\varphi_{ki} = 45°$ and (c.II)) $\varphi_{ki} = 90°$. (d.I) $V/V_{max}$ as a function of $\varphi_H$ for $\varphi_{ki}$ fixed at 0° and magnetic fields of 300 Oe and 2000 Oe. (d.II) $V/V_{max}$ as a function of $\varphi_H$ for several $\varphi_{ki}$ values at a magnetic field of 300 Oe.

III. COMPARISON WITH EXPERIMENTAL RESULTS

Three magnetic systems with distinct anisotropy configurations were produced to test the robustness of the theoretical approach.

The first sample is a YIG (3 m)/Pt (6 nm) bilayer grown onto a [111]-oriented $Gd_3Ga_5O_{12}$ substrate, a sample with isotropic magnetic properties. The sample geometry and the employed materials make this sample suitable to the study of pure LSSE. The second sample is a $Co_{40}Fe_{40}B_{20}$ (300 nm) film grown onto a glass substrate, corresponding to a uniaxial magnetic system. In this case, given the sample structure, the thermoelectric voltage is associate with the pure ANE response. The third sample consists in a $Co_{58.73}Fe_{27.83}Al_{13.44}$ (53 nm)/W (2 nm) bilayer deposited onto [100]-oriented GaAs substrate. Such system has cubic magnetocrystalline and uniaxial magnetic anisotropies. Here, the thermoelectric response is a result of the mixing of LSSE and ANE. Although the LSSE and ANE have different physical origins, their correspondence with respect to hysteretic behavior is similar. The experimental procedures used for sample preparation and characterization are presented in the Supplementary Material.

The investigated samples allow verifying not just the dependence of the magnetic and thermoelectric features with the magnetic anisotropy, but they also provide tools to show that the ANE and LSSE experiments present very similar behavior and can be described by the developed theoretical approach.

Figure 7 shows the experimental and theoretical results for the YIG (3 m)/Pt (6 nm) bilayer. The measurements are obtained with $\Delta T = 27$ K. For the calculations, it is considered $m_s = 143$ emu/cm³, $\lambda_{eff} = 5.2 \times 10^{-5}$ V/K, and $h_{ki} = 4.8$ Oe. Further, it is assumed $\varphi_{ki} = 30° + \varphi_H$. This value allows to reproduce the experimental results, for which it is observed a deviation between $\varphi_k$ and $\varphi_H$. Anyway, the isotropic behavior of the YIG (3 m)/Pt (6 nm) bilayer as a function of the

$\varphi_H$ is evident and corroborates the decrease of $V/V_{max}$ with increasing $\varphi_H$ and the absence of modifications in the shape of the curve with varying field orientation. Considering the curves obtained at $\varphi_H = 90°$ (not show), no thermoelectrical signal is observed. This feature is associated with the perpendicular alignment of the electric field and the electric contact direction (see Fig. 1).

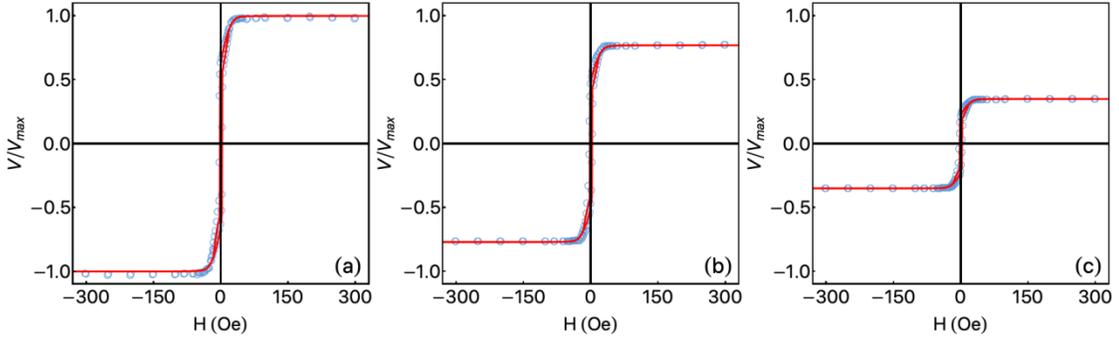

FIG. 7. **Comparison between experimental and theoretical results.** Normalized magnetization curves obtained for (a) $\varphi_H = 0°$, (b) $\varphi_H = 30°$. (c) $\varphi_H = 60°$. The experimental results are obtained for the YIG (300 nm)/Pt (6 nm) bilayer grown onto [111]-oriented Gd$_3$Ga$_5$O$_{12}$ substrate. The open circles depict the experimental results, while the red line brings the numerical calculations obtained with the theoretical approach for an isotropic magnetic system. To perform the calculation, a system with $m_s = 143$ emu/cm³, $h_{ki} = 4.8$ Oe, $\varphi_{ki} = 30° + \varphi_H$, and $\lambda_{eff} = 5.2 \times 10^{-5}$ V/K is considered. Besides, $\Delta T = 27$ K has been used during the LSSE measurement.

Considering the uniaxial magnetic system, the experimental results for a Co$_{40}$Fe$_{40}$B$_{20}$ film deposited onto a glass substrate are presented. During the deposition a magnetic field of 1.0 kOe has been applied to induce the uniaxial magnetic behavior. Figure 8 presents thermoelectric curves for the selected $\varphi_H$ values of 0°, 30°, and 60°. For the numerical calculations, $m_s = 1026$ em/cm³, $k_{ui} = 7.072 \times 10^4$ ergs/cm³, $k_{c1} = 0$, $\Delta T = 27$ K, $\lambda_{eff} = 1.3 \times 10^{-5}$ V/K, and $\varphi_{ki} = 85°$ have been used

From figure 8, a very interesting evolution in the shape of the thermoelectric response with $\varphi_H$ is observed. In particular, this feature is a consequence of the change in the orientation of the field, which is rotating from the easy magnetization axis direction to the hard magnetization one. For $\varphi_H = 0°$, i.e. along the easy axis, it is verified a mirror between the thermoelectric and magnetization curves (the latter not shown). This behavior is due to the fact that the detection direction of the thermoelectric voltage, defined by the electrical contacts, is perpendicular to the easy magnetization axis. By rotating the magnetic field to $\varphi_H = 30°$ and $\varphi_H = 60°$, at a high magnetic field regime, the component of the induced electric field along the direction of the electrical contacts decreases, leading a reduction of the thermoelectrical voltage. On the other hand, at a low magnetic field regime, the uniaxial magnetic anisotropy leads the magnetization to the easy axis, rotating the electric field and, consequently, increasing the thermoelectrical voltage, as confirmed from the results. This mechanism is observed in both, experimental results and numerical calculations. Here, it is worth pointing out the alignment between the electric field and contact direction is essential to describe the experimental curves. In addition, although a multidomain system is simulated, the complex domain structure of a real system leads to an increase in the coercive field for a real sample. This feature is responsible for the small discrepancy between the coercivity values of the experimental and theoretical results observed as $\varphi_H$ increases.

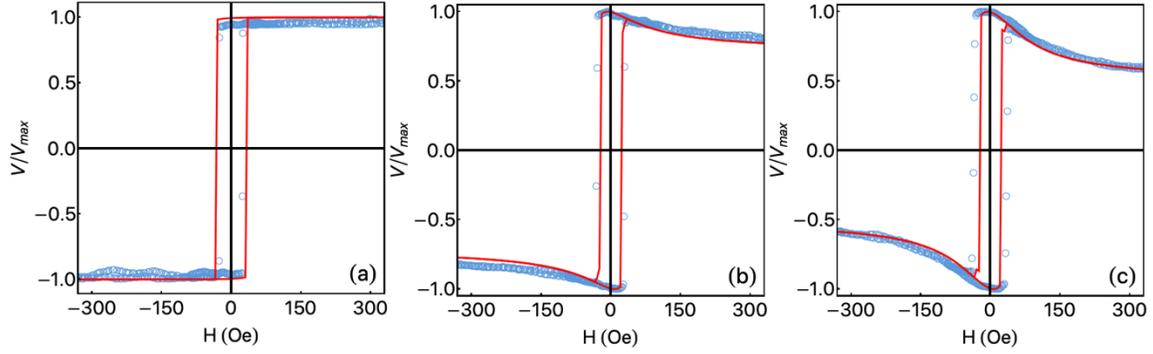

FIG. 8. **Comparison between experimental and theoretical results.** Normalized magnetization curves obtained for (a) $\varphi_H = 0°$, (b) $\varphi_H = 30°$. (c) $\varphi_H = 60°$. The experimental results are obtained for the $Co_{40}Fe_{40}B_{20}$ film grown onto the glass substrate, presenting uniaxial magnetic anisotropy behavior. The open circles depict the experimental results, while the red line brings the numerical calculations obtained with the theoretical approach for an isotropic magnetic system. To perform the calculation, it has been considered $m_s = 1026$ em/cm³, $k_{ui} = 7.072 \times 10^4$ ergs/cm³, $k_{c1} = 0$, $\Delta T = 27$ K, $\lambda_{eff} = 1.3 \times 10^{-5}$ V/K and $\varphi_{ki} = 85°$.

Figure 9 shows the experimental and theoretical results obtained for the thermoelectric curves of $Co_{58.73}Fe_{27.83}Al_{13.44}$/W bilayers grown onto [100]-oriented GaAs substrates and submitted to a 1.0 kOe during the deposition (details on the experimental procedure can be found in Supplementary Material). The employed substrates and the annealing process allows to obtain a magnetic system with uniaxial and cubic magnetocrystalline anisotropies. For the numerical calculations, it has been considered $m_s = 1330$ emu/cm³, $k_{ui} = 1.66 \times 10^5$ ergs/cm³, $k_{c1i} = 2.8\, k_{ui}$, $\Delta T = 27$ K, an $\varphi_{ki} = 90°$, and $\lambda_{eff} = 3.79 \times 10^{-8}$ V/K. From a general point of view, the evolution of the curve's shape as $\varphi_H$ increases has been observed. The curves present a remarkable modification on the shape when $\varphi_H$ is higher than 45°. This behavior is similar to the one observed in the theoretical results presented in Fig. 6. As discussed before, at $\varphi_k = 45°$, the studied system shows an intermediate magnetization axis. For this reason, the $\varphi_H$ value is an important angle. For $\varphi_H$ values smaller than this critical value, the thermoelectric curves seem to present similar behavior to the one presented by the magnetization. On the other hand, for $\varphi_H$ values larger than 45°, the shape changes drastically. Hence, the cross of $\varphi_H$ through this anisotropy axis leads to modifications in the shape of the thermoelectric response, an issue associated with the anisotropy configuration, discussed in detail in Ref. [23].

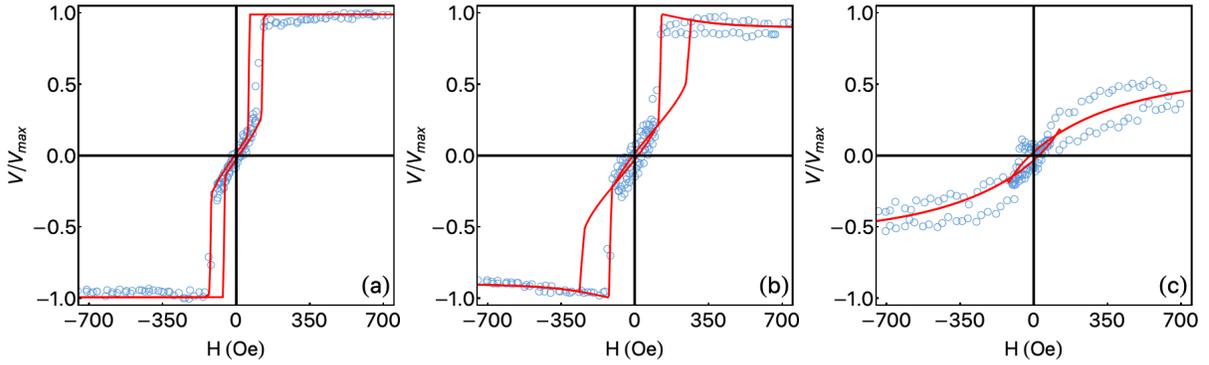

FIG. 9. **Comparison between experimental and theoretical results.** Normalized magnetization curves obtained for (a) $\varphi_H = 0°$, (b) $\varphi_H = 30°$. (c) $\varphi_H = 60°$. The experimental results are obtained for the Co$_{58.73}$Fe$_{27.83}$Al$_{13.44}$/W bilayer grown onto [100]-oriented GaAs substrate, a system having cubic and uniaxial magnetic anisotropies. The open circles depict the experimental results, while the red line brings the numerical calculations obtained with the theoretical approach for an isotropic magnetic system. To perform the calculation, it has been applied $m_s = 1330$ emu/cm³, $k_{ui} = 1.66 \times 10^5$ ergs/cm³, $k_{c1} = 2.8 k_{ui}$, $\Delta T = 27$ K, and $\varphi_{ki} = 90°$, $\lambda_{eff} = 3.79 \times 10^{-8}$ V/K.

## IV. CONCLUSION

In summary, the role of magnetic anisotropies on the magnetic and thermoelectric responses of noninteracting multidomain ferromagnetic systems has been addressed, based on a theoretical approach based in a modified Stoner Wohlfarth model. The approach has been tested for an isotropic system, a uniaxial magnetic one, as well as for a system having a mixture of uniaxial and cubic magnetocrystalline magnetic anisotropies. It has been verified remarkable modifications of the magnetic behavior with the anisotropy and it has been shown that the thermoelectric response is strongly affected by these changes. Further, the fingerprints of the energy contributions to the thermoelectric response have been disclosed, revealing that the anisotropy energy terms represent a fundamental contribution to the energy balance even in saturated states when the field is right above the saturation field. To test the robustness of the developed theoretical approach, three films have been engineered with specific magnetic properties and have been compared to the corresponding magnetic and thermoelectric experimental data. From the comparison and the remarkable agreement between experiment and theory, experimental evidence has been provided to confirm the validity of the theoretical approach. The results go beyond the traditional reports focusing on magnetically saturated films and show how the thermoelectric effect behaves during the whole magnetization curve. The results emphasize the importance of selecting specific field alignments and field amplitudes to provide reliable interpretation of the ANE and LSSE experiments. Our findings reveal a suitable and robust way to explore the ANE and LSSE effects as a powerful tool to study magnetic anisotropies, as well as to employ systems with magnetic anisotropy as sensing elements for technological applications.

ACKNOWLEDGMENTS

This work was partially supported by the Brazilian agencies CNPq and CAPES. M. A. Correa acknowledges CNPq through the project 304943/2020-7, and CAPES through process 88887.573100/2020-00. Further, this work was also supported by the Portuguese Foundation for Science and Technology (FCT) in the framework of the Strategic Funding UID/FIS/04650/2019 and project PTDC/BTM-MAT/28237/2017. A. Ferreira acknowledges the FCT for the Junior Research

Contract (CTTI-31/18). Financial support from the Basque Government Industry Department under the ELKARTEK, and PIBA (PIBA-2018-06) programs is also acknowledged.